\documentclass{iaus}
\usepackage{graphicx}

\title[S0 Galaxies in Fornax] %% give here short title %%
{Central Stellar Populations of S0 Galaxies in the Fornax Cluster}

\author[Bedregal et al.]   %% give here short author list %%
{A.G. Bedregal$^{1}$, A. Arag\'on-Salamanca$^{1}$, M.R. Merrifield$^{1}$ and N. Cardiel$^{2}$}
\affiliation{$^{1}$School of Physics and Astronomy, Univ.\ of Nottingham,
  Nottingham, NG7 2RD, U.K. \\[\affilskip]$^{2}$Departamento de Astrof\'isica,
  Facultad de F\'isicas, Univ. Complutense de Madrid, Spain}
\pubyear{2007}
\volume{241}  %% insert here IAU Symposium No.
\pagerange{xxx--xxx}
\date{?? and in revised form ??}
\jname{Proceedings Title IAU Symposium}
\editors{A.C. Editor, B.D. Editor \& C.E. Editor, eds.}
\begin{document}

\maketitle

\begin{abstract}
 Based on FORS2-VLT long-slit spectroscopy, the analysis of the central
   absorption line indices of 9 S0 galaxies in the Fornax Cluster is
   presented. Central indices correlate with central velocity dispersions ($\sigma_0$)
   as observed in ellipticals (E). However, the stellar population properties of
   these S0s indicates that the observed trends are produced by
   relative differences in age and $\alpha$-element abundances and not in
   metallicity ([Fe/H])
   as previous studies have found in E galaxies. The observed scatter in the
   line indices versus $\sigma_0$ relations can be partially explained by the
   rotationally-supported nature of many of these systems. The presence of
   tighter line indices vs.\ maximum (circular) rotational velocity ($V_{\rm MAX}$) relations
   confirms this statement. It was also confirmed that the dynamical mass is
   the driving physical property of all these correlations and in our Fornax
   S0s it has to be estimated assuming rotational support.
\end{abstract}

\firstsection % if your document starts with a section,

\section{The Index$^*$ vs.\ $\log(\sigma_0)$ Relations}

% NOTE use of \upi in above paragraph and subsequently throughout paper.
% The Greek constant character pi should be upright.

   In Figure~1 (left), different central line indices are plotted against $\sigma_0$ for our
   sample (Bedregal et al.\ 2006). Bright and faint S0s lay in two separate clumps, each one in
   opposite extremes of the $\sigma_0$ range. When models from \cite{BC03} are used to estimate stellar population parameters, the observed trends seem to be driven by age and $\alpha$-element relative abundances. This is opposite to what is usually found in E, where [Fe/H] is an important driver of the correlations. By parametrising line indices using \cite{BC03} models we found that age can explain the observed slopes of H$\beta$-log($\sigma_0$) and Fe-log($\sigma_0$) relations, while additional differences in $\alpha$-element relative abundances would be required to explain the Mg-log($\sigma_0$) trends. On the other hand, [Fe/H] by itself cannot explain any of the observed Index$^*$-log($\sigma_0$) relations.   

\begin{figure} 
\includegraphics[scale=0.31]{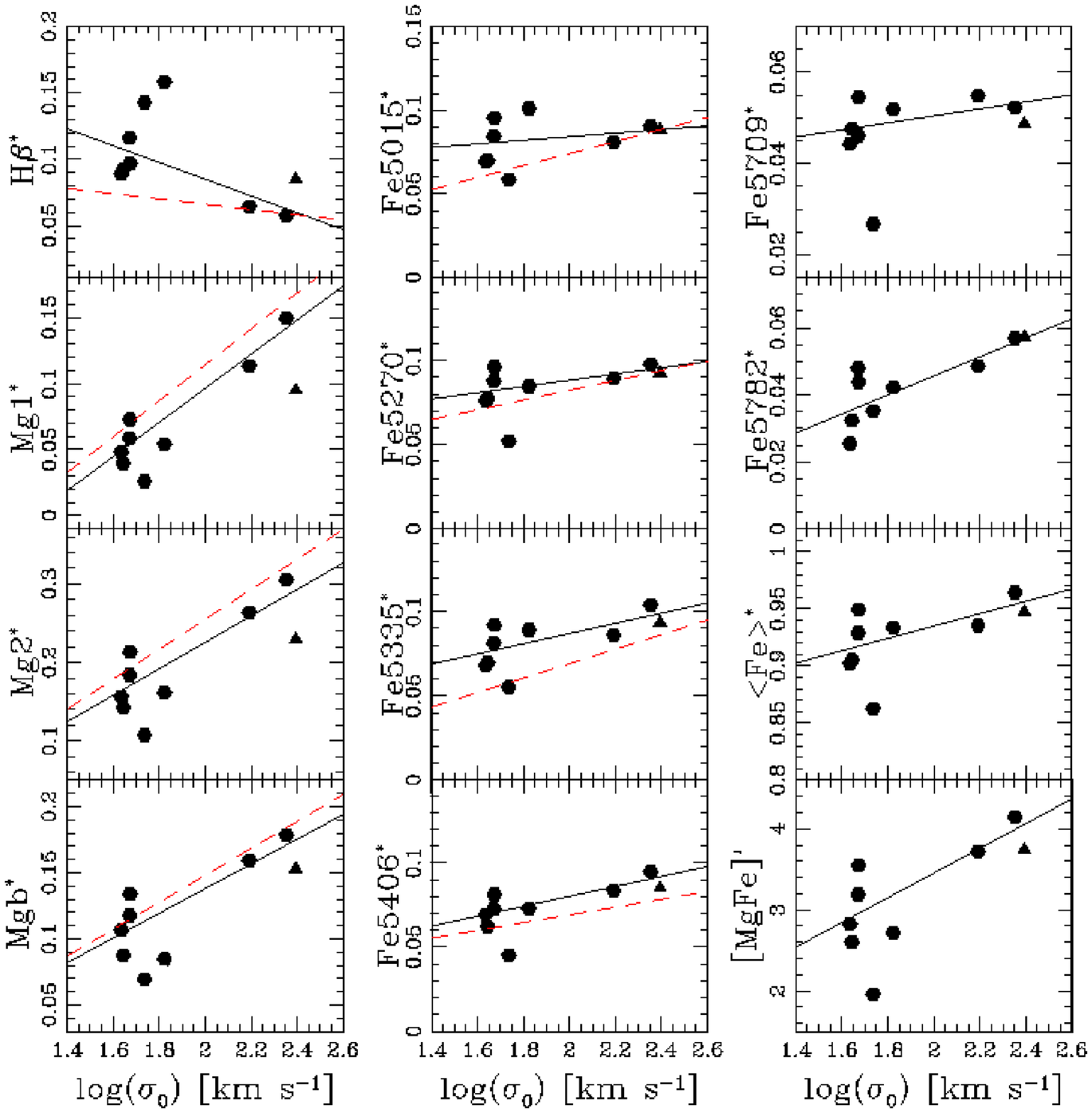}
\includegraphics[scale=0.31]{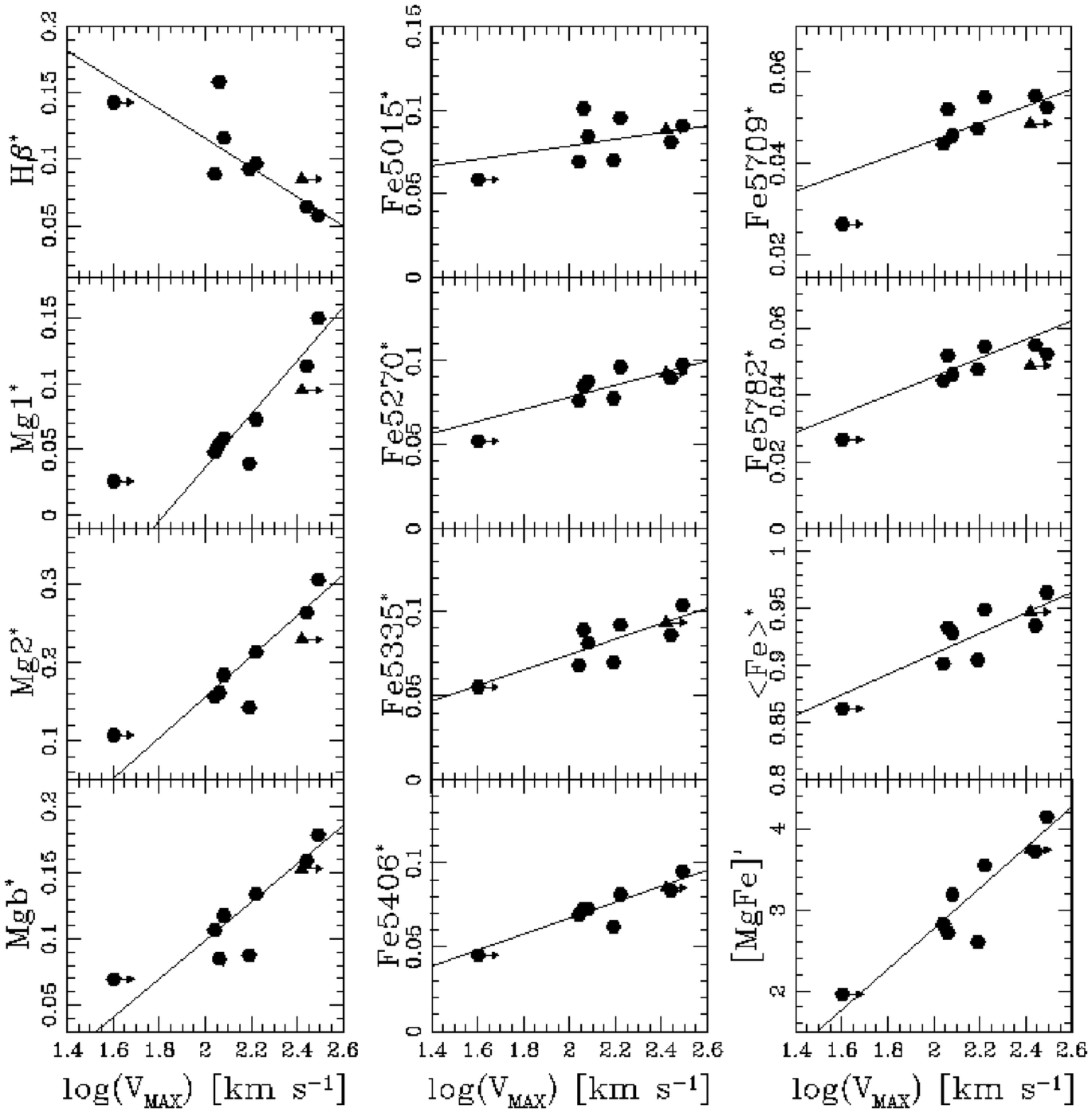}
\caption{Line indices vs.\ $\sigma_0$ (left) and $V_{\rm
    MAX}$ (right) for 9 S0s in the Fornax Cluster. Continuous lines: best fits
    to the data. Dashed lines: best fits found by \cite{K00} for Es in
    Fornax. Triangle: Fornax A, a merger remnant. Two dots with arrows:
    pressure supported systems, for which the deprojected azimuthal velocity
    was used as a lower limit of $V_{\rm MAX}$.}
\end{figure}

\section{The Index$^*$ vs. $\log(V_{\rm MAX})$ Relations:} 
   In Figure~1 (right) we present the Index$^*$ vs.\ log($V_{\rm MAX}$)
   relations for our sample of S0s. Clear trends appear in all the panels. It
   is interesting to notice that the standard deviation of the linear fits
   performed is smaller for almost all the indices in comparison to its
   counterpart from the Index$^*$-log($\sigma_0$) fits. Assuming that mass is
   the fundamental physical parameter governing the properties of these
   galaxies, the improvement in the fits may be interpreted in the sense that
   $V_{\rm MAX}$ is a better estimator of the dynamical mass than
   $\sigma_0$. To test this hypothesis, the mass of these galaxies was
   parametrised as a function of $\sigma$ and $V_{\rm MAX}$ in order to
   compare how central indices correlate with both possible
   descriptions. $R_{\rm d}$ and $R_{\rm e}$ are the disk scale and effective
   radius of the bulge respectivelly and the mean $\sigma$ was estimated
   within $R_{\rm e}$ (see Figure~2).

\begin{figure} 
\includegraphics[scale=0.31]{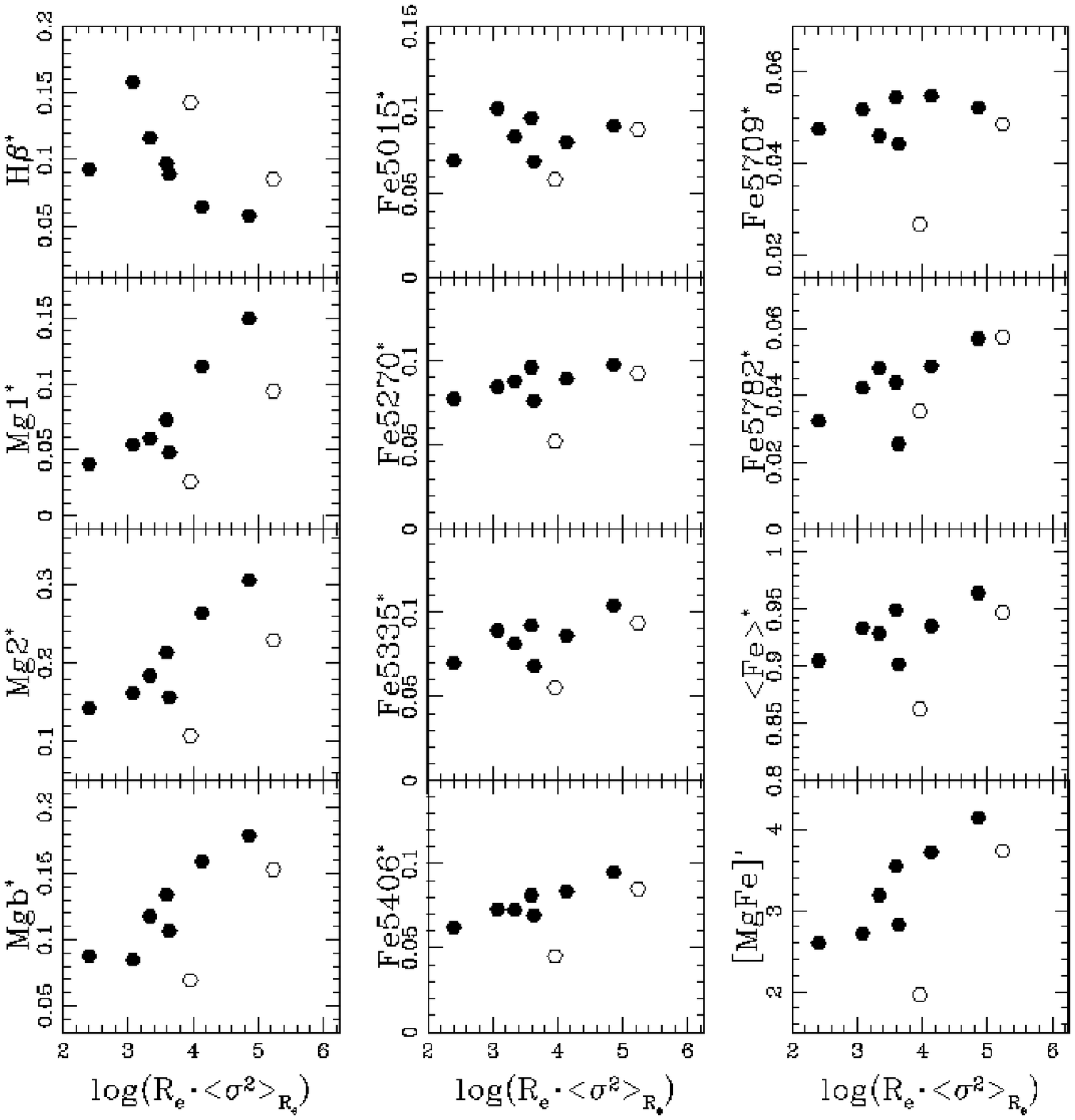}
\includegraphics[scale=0.31]{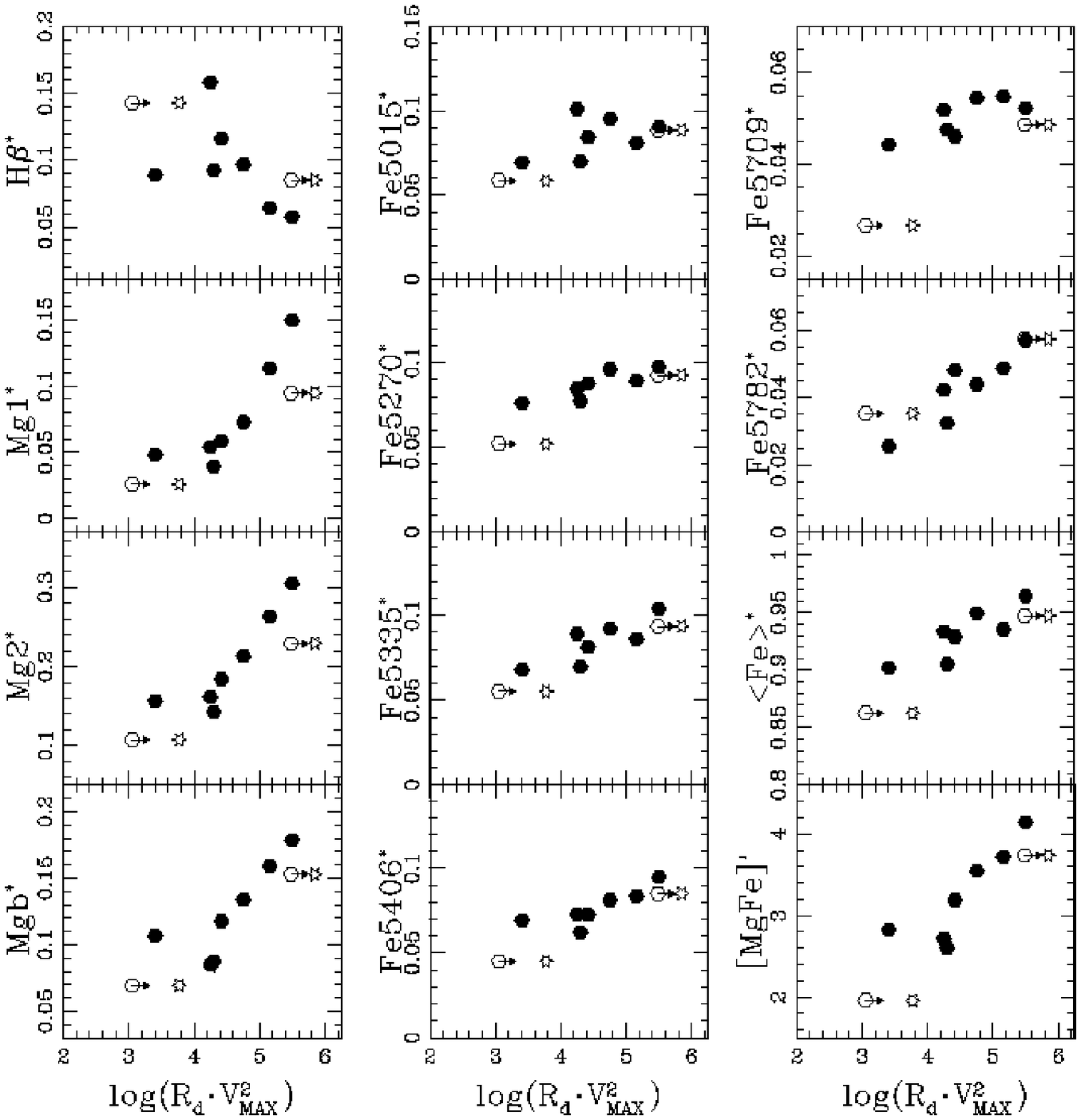}
\caption{ Line indices vs.\ dynamical mass estimated from $\sigma$ (left) and
  from $V_{\rm MAX}$ (right). Open symbols: two pressure supported
  systems (arrows as Fig~1). Open stars: dynamical mass of these two galaxies estimated from
  $\sigma$ by assuming an isothermal sphere in hydrostatic equilibrium.}
\end{figure} 

Clearly, both $\sigma$ and $V_{\rm MAX}$  are tracing the mass of these systems. However, when
mass is estimated using $\sigma$ (Figure~2, left panel), it is underestimated in rotationally
supported systems with respect to pressure supported ones. On the other hand,
the right panel of Figure~2 shows that galaxies follow a common sequence when $V_{\rm MAX}$ is used. The
tighter trends in this figure agree with the idea that $V_{\rm MAX}$ is a better tracer
of dynamical mass than $\sigma$ for this sample of galaxies.

\end{document}